\documentclass[%
 reprint,
 amsmath,amssymb,
 aps,
]{revtex4-2}

\usepackage{graphicx}
\usepackage{dcolumn}
\usepackage{bm}
\usepackage{here}
\usepackage{enumerate}

\makeatletter
\let\MYcaption\@makecaption
\makeatother
\usepackage{subcaption}
\makeatletter
\let\@makecaption\MYcaption
\makeatother

\usepackage{hyperref}

\begin{document}


\title{Localization of gravitational waves using machine learning}

\author{Seiya Sasaoka}
\affiliation{Department of Physics, Tokyo Institute of Technology, 2-12-1 Oh-okayama, Meguro, Tokyo, 152-8551 JAPAN}%

\author{Hirotaka Takahashi}
\affiliation{Research Center for Space Science, Advanced Research Laboratories, Tokyo City University, 8-15-1 Todoroki, Setagaya, Tokyo, 158-0082 JAPAN}%

\author{Yilun Hou}
\affiliation{Department of Physics, Tokyo Institute of Technology, 2-12-1 Oh-okayama, Meguro, Tokyo, 152-8551 JAPAN}%

\author{Kentaro Somiya}
\affiliation{Department of Physics, Tokyo Institute of Technology, 2-12-1 Oh-okayama, Meguro, Tokyo, 152-8551 JAPAN}%

\date{\today}

\begin{abstract}
An observation of gravitational waves is a trigger of the multi-messenger search of an astronomical event. A combination of the data from two or three gravitational wave telescopes indicates the location of a source and low-latency data analysis is key to transferring the information to other telescopes sensitive at different wavelengths. In contrast to the current method, which relies on the matched-filtering technique, we proposed the use of machine learning that is much faster and possibly more accurate than matched filtering. 
Our machine-learning method is a combination of the method proposed by Chatterjee {\it et al.} and a method using the temporal convolutional network.
We demonstrate the sky localization of a gravitational-wave source using four telescopes: LIGO H1, LIGO L1, Virgo, and KAGRA, and compare the result in the case without KAGRA to examine the positive influence of having the fourth telescope in the global gravitational-wave network.
\end{abstract}


\maketitle

\section{Introduction\label{sec:intro}}
The Laser Interferometer Gravitational-wave Observatory (LIGO)~\cite{ref:aasi2015advanced} located in Livingston, Louisiana (L1) and Hanford, Washington (H1) in the USA made the first observation of a gravitational wave (GW) from binary black holes in September 2015~\cite{ref:Abbott2016}. Since then, LIGO and Virgo~\cite{ref:acernese2014advanced} in Europe have conducted three international joint observation runs, which included the observation of 90 events of GWs emitted by the coalescence of compact binaries~\cite{ref:Abbott2019,ref:Abbott2020,ref:Abbott2021_a,ref:Abbott2021_b}. In addition, GEO600~\cite{ref:grote2008status} in Germany and KAGRA~\cite{ref:akutsu2018kagra} in Japan conducted a two-week joint observation run (O3GK) in April 2020. The fourth international joint observation run will be conducted by LIGO, Virgo, and KAGRA.

Progress in information science, particularly in the fields of artificial intelligence, machine learning, and deep learning, has been remarkable. We can expect many benefits by applying machine learning (including deep learning) to gravitational-wave data analysis. 
The application of knowledge from other fields (e.g., audio processing, image processing, and processing of biological signals such as electrocardiogram, electromyogram, and electroencephalogram) to gravitational-wave data analysis is considered key to realizing the potential of multi-messenger astronomy.

GWs from the coalescence of compact binaries are usually analyzed using matched filtering, which uses model waveforms. However, the computational cost is often a problem. As an alternative, machine learning is increasingly being applied in the analysis of various gravitational-wave
data after pioneering work by George and Huerta~\cite{ref:George_2018}. For example, methods have been proposed to estimate parameters for GW signals emitted from the coalescence of compact binaries~\cite{ref:Chua_2020,ref:Gabbard_21}. Another example concerns core-collapse supernovae. It is generally difficult to create an appropriate model waveform for GWs from core-collapse supernovae, rendering matched filtering unsuitable. In view of this, machine learning has been introduced as an alternative~\cite{ref:Astone_2018,ref:Iess_2020,ref:Chan_2020,ref:Lopez_2021}. Research is also underway as a method to analyze continuous waves~\cite{ref:Dreissigacker_2019,ref:Dreissigacker_2020,ref:Beheshtipour_2020,ref:Beheshtipour_2021,ref:Yamamoto_2021}. The Gravity Spy project~\cite{ref:Zevin_2017,ref:Bahaadini_2018,ref:Soni_2021,ref:Bahaadini_2017} and \citeauthor{ref:sakai_2021}~\cite{ref:sakai_2021} have also used machine learning to classify the transient noises.
Furthermore, machine learning, specifically deep learning, has been used for waveform modeling~\cite{ref:Alvin_2021}. 

We focused on the localization of the GWs from a binary black hole. Localization requires determining the right ascension and declination of an event. It is important for multi-messenger astronomy to estimate these parameters accurately and rapidly when processing GW signals. One way to use deep learning to localize GW signals is to treat the problem as a classification task by dividing the sky into several sectors.  \citeauthor{Chatterjee2019}~\cite{Chatterjee2019} extracted seven features from the strain data of LIGO H1, LIGO L1, and Virgo, used them as input to an artificial neural network (ANN), and then trained the ANN to classify GW signals from binary black hole mergers. They tested the model using the samples with parameters of real events and showed that the model could localize each signal with high accuracy in a much shorter time than the current localization technique.

In this research, we considered the following three methods to classify GW signals from binary black hole mergers for sky localization: (i) the method proposed by \citeauthor{Chatterjee2019}~\cite{Chatterjee2019}, (ii) a temporal convolutional network (TCN) \cite{TCN}, and (iii) a combination of (i) and (ii).
These methods were then applied to the simulated data from LIGO and Virgo to compare the localization accuracy. Finally, we trained the model using simulated strain data from all the four telescopes, including KAGRA, to compare the results with the data from the three telescopes.

The remainder of this paper is organized as follows. Section~\ref{sec:methods} describes our datasets and the details of the three methods. Evaluations of each model and the results of the localization using four telescopes are presented in Sec.~\ref{sec:results}. We summarize and conclude the paper in Sec.~\ref{sec:conclusions}.

\section{Set up of sky localization analysis\label{sec:methods}}
%
%
\begin{figure*}[tbh]
  \begin{minipage}{0.31\linewidth}
    \centering
    \includegraphics{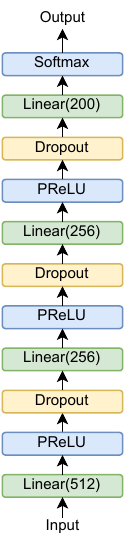}
    \subcaption{}
    \label{fig:ANN}
  \end{minipage}
  \begin{minipage}{0.31\linewidth}
    \centering
    \includegraphics{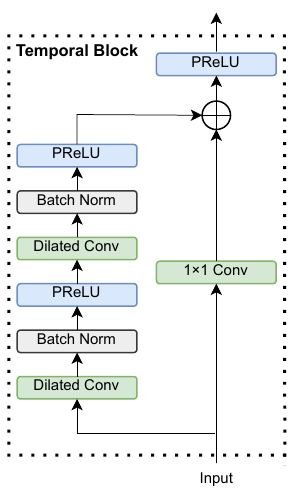}
    \vspace{0.75cm}
    \subcaption{}
    \label{fig:temporal_block}
  \end{minipage}
  \begin{minipage}{0.31\linewidth}
    \centering
    \includegraphics{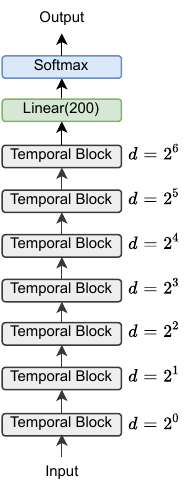}
    \hspace{-1.2cm}\vspace{0.85cm}
    \subcaption{}
    \label{fig:TCN}
  \end{minipage}
  \caption{The structure of our models. (a) ANN architecture for method I. (b) Temporal block in a TCN. (c) TCN architecture for method II.}
\end{figure*}
%
%
In this section, we describe the data generation process, details of the three methods, and the training process.
\subsection{Data generation process}
We used ggwd~\cite{ggwd}, an open-source tool developed by \citeauthor{Gebhard2019}~\cite{Gebhard2019}, to generate the training and test data. 
This tool originally generated GW strain data from LIGO H1 and LIGO L1, but we edited the code so that it could also generate data from Virgo and KAGRA. 
The waveforms were simulated in the time domain using the \verb|SEOBNRv4| model~\cite{SEOBNRv4} with a sampling rate of 2048 Hz, and cut to a length of 0.25 seconds. The waveform parameters used are listed in Table~\ref{table:parameter}. The signals are scaled in order that the samples have the desired signal to noise ratio (SNR), then Gaussian noise was added. This noise was simulated using the power spectral density of each detector. The power spectral density of the advanced LIGO detectors was obtained from their document server~\cite{LIGO-T1800044-v5}, that of advanced Virgo is the one shown in Ref.~\cite{ref:acernese2014advanced}, and that of KAGRA was obtained from their document server~\cite{JGW-T1707038-v9}. The resulting samples were whitened and filtered using a bandpass filter to select frequencies between 18 Hz and 500 Hz. We generated 200,000 samples for training and 40,000 samples for validation and testing.

\subsection{Method I: ANN}
First, to treat the problem as a classification task, we divided declination and right ascension by the same angle and labeled each sector. 
In particular, whenever we divided the declination into three, the right ascension was divided into six, resulting in 18 sectors. The number of sectors considered in this study ranged from 18 to 200. We used the three methods (ANN, TCN, and combined) to classify each GW signal into one of the sectors. 

\begin{table}[tbh]
  \caption{Parameters used to simulate the signals.}
  \label{table:parameter}
  \centering
  \begin{tabular}{lcr}
    \hline
     Mass1, Mass2 & [30M$_\odot$, 80M$_\odot$]    \\
     Spin1z, Spin2z & [0, 0.998]  \\
     Right ascension & [0, $2\pi$]    \\
     Declination & [$-\pi/2$, $\pi/2$]   \\
     Coalescence phase & [0, $2\pi$]    \\
     Inclination & [0, $\pi$] \\
     Polarization & [0, $2\pi$] \\
     SNR & [10, 50]\\
    \hline
  \end{tabular}
\end{table}

For method I, we use the method proposed by \citeauthor{Chatterjee2019}~\cite{Chatterjee2019}. First, the following seven features were computed, then entered as input into the network: (i) arrival time delays of signals; (ii) maximum cross-correlation values of signals; (iii) arrival time delays of analytic signals; (iv) maximum cross-correlation values of analytic signals; (v) ratios of amplitudes
around merger; (vi) phase lags around the merger; and (vii) complex correlation coefficients of signals. Because all of these features were computed using two sets of strain data, when using three (or four) telescopes for localization, we had 21 (or 42) input features. Each feature was then standardized to have a mean of zero and a standard deviation of one. The structure of the resulting ANN model used in this study is shown in Fig.~\ref{fig:ANN}. The ANN structure was modified from that used in the previous study~\cite{Chatterjee2019} to realize a similar or a slightly better localization accuracy.

\subsection{Method II: TCN}
For method II, the raw time-series data were input to the TCN. A TCN is a model used for sequential-data modeling and consists of several temporal blocks (Fig.~\ref{fig:temporal_block}) that have dilated causal convolutions and residual connections. The structure of the network is shown in Fig.~\ref{fig:TCN}. We stacked seven temporal blocks with the dilation $d=2^i$ $(i=0, \cdots, 6)$ and kernel size 3 for  the $i$-th block. Because the TCN is used for classification and not for forecasting, the causal structure of the dilated convolutional layer was removed. The input had three (or four) channels when using strain data from three (or four) telescopes, and the number of output channels was 64 for each temporal block. The last linear layer output the tensor with the size of the number of sectors. The parameters in the network were initialized by He's method~\cite{He}.

\subsection{Method III: Combined}
Method III is a combination of methods I and II. There are several ways to combine the two deep learning models. A naive way would be to compare the output predictions with the two methods and select the sector with the larger prediction. Here, we take the weighted average of the output predictions and select the sector that provides better accuracy. The weight is decided to maximize the accuracy of the validation data.

\subsection{Training process}
Each deep learning model in methods I and II was implemented using Pytorch~\cite{Pytorch}. For training the models, cross entropy was used as the loss function and Adam~\cite{adam} was used to optimize the weights. The initial learning rate was set to 0.001, but during training, it was controlled using Pytorch’s \verb|ReduceLROnPlateau|. We trained the ANN for 200 epochs and TCN for 25 epochs with mini-batch sizes of 4000 and 256, respectively. Meanwhile, we only used the checkpoints for each model with the lowest validation loss for the test.

\section{Results\label{sec:results}}
\begin{figure*}[htbp]
    \begin{tabular}{cc}
      \begin{minipage}[t]{0.45\hsize}
        \centering
        \includegraphics[keepaspectratio,scale=0.4]{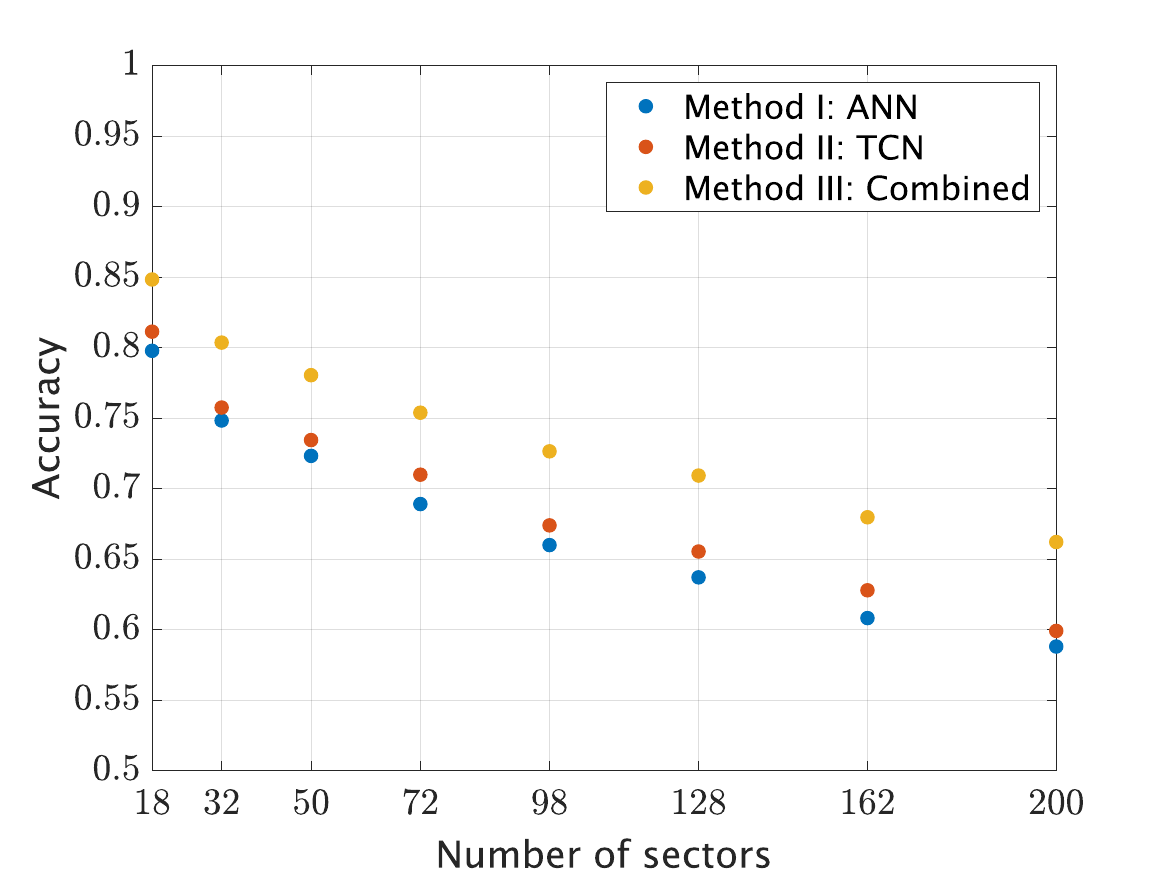}
        \caption{\label{fig:method123}The localization accuracy using three telescopes.}

      \end{minipage} &
      \begin{minipage}[t]{0.45\hsize}
        \centering
        \includegraphics[keepaspectratio, scale=0.4]{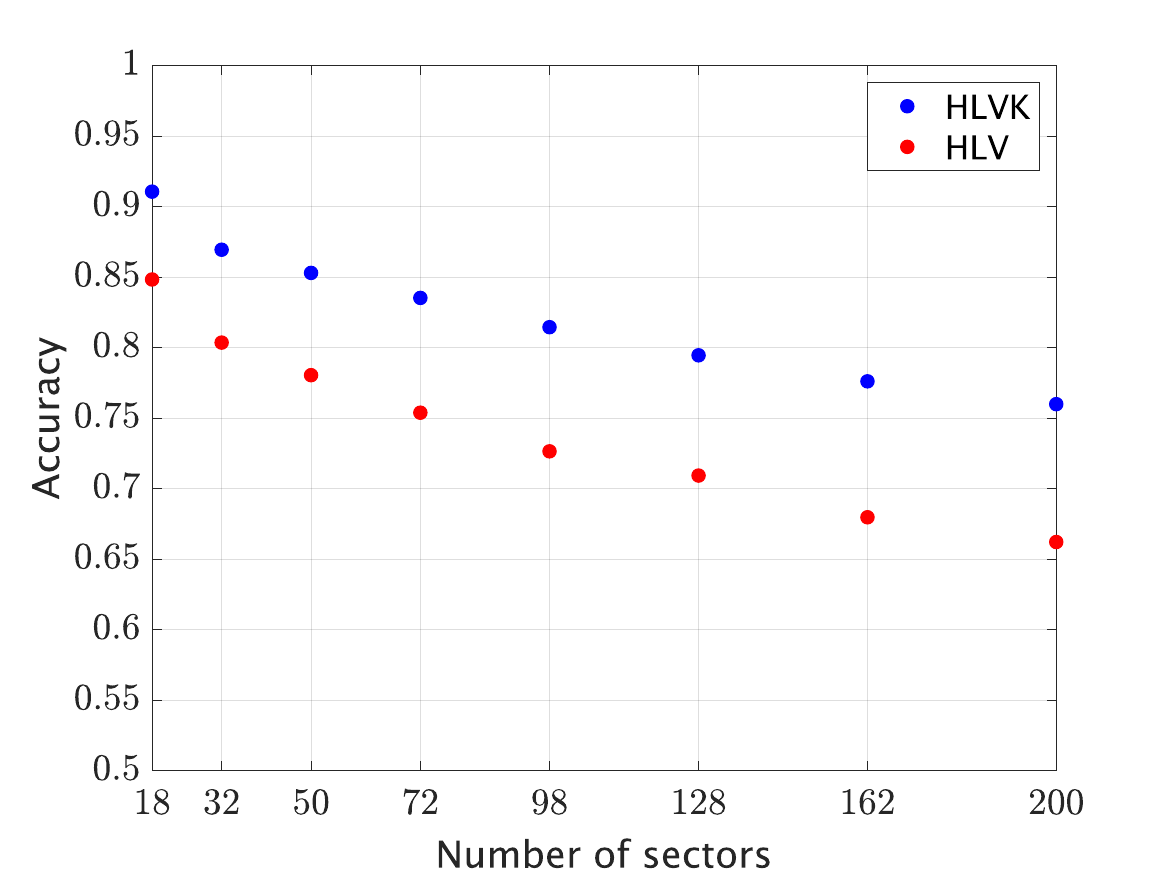}
        \caption{\label{fig:HLVK} The localization accuracy using three and four telescopes. HLV means using LIGO H1, L1, and Virgo. HLVK means using LIGO H1, L1, Virgo, and KAGRA. } 
      \end{minipage} \\
      \begin{minipage}[t]{0.45\hsize}
        \centering
        \includegraphics[keepaspectratio, scale=0.4]{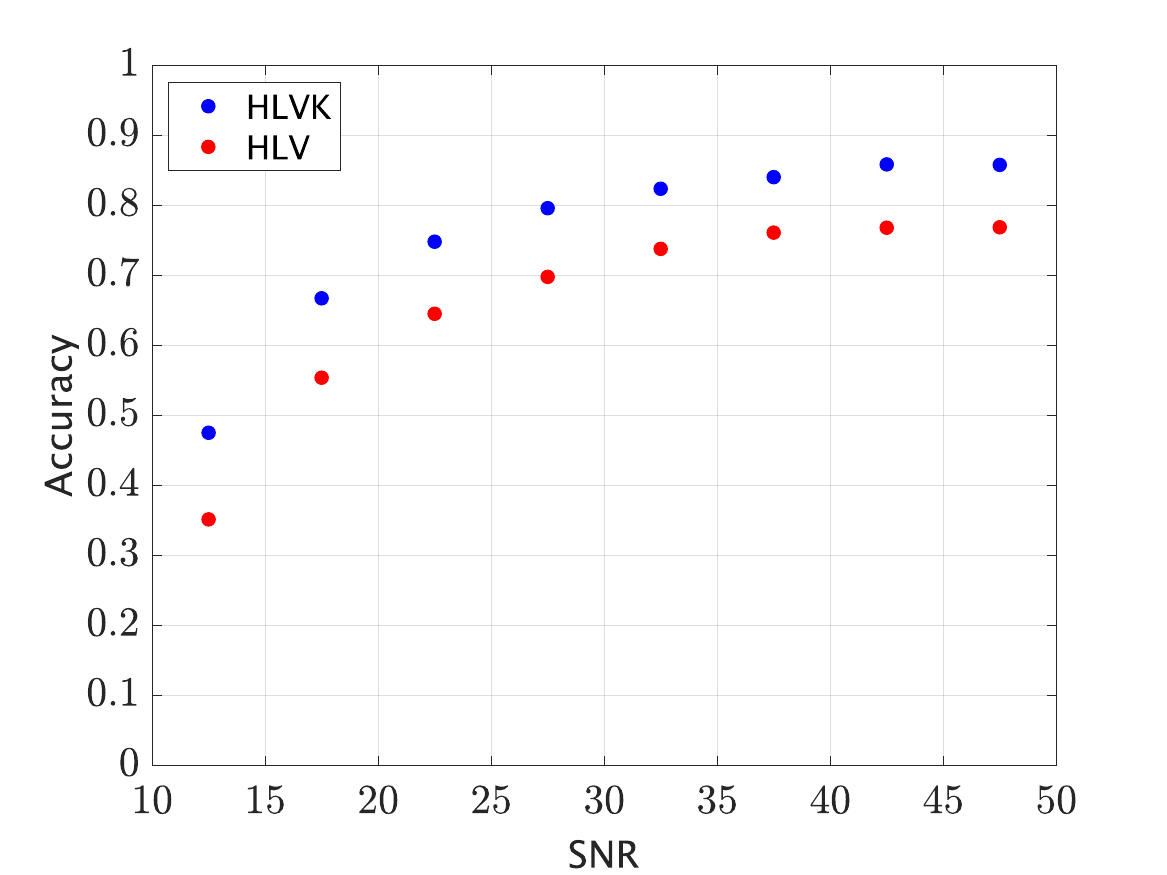}
       \caption{\label{fig:acc_snr} The localization accuracy for 200 sectors vs SNR.} 
      \end{minipage} &
      \begin{minipage}[t]{0.45\hsize}
        \centering
        \includegraphics[keepaspectratio, scale=0.4]{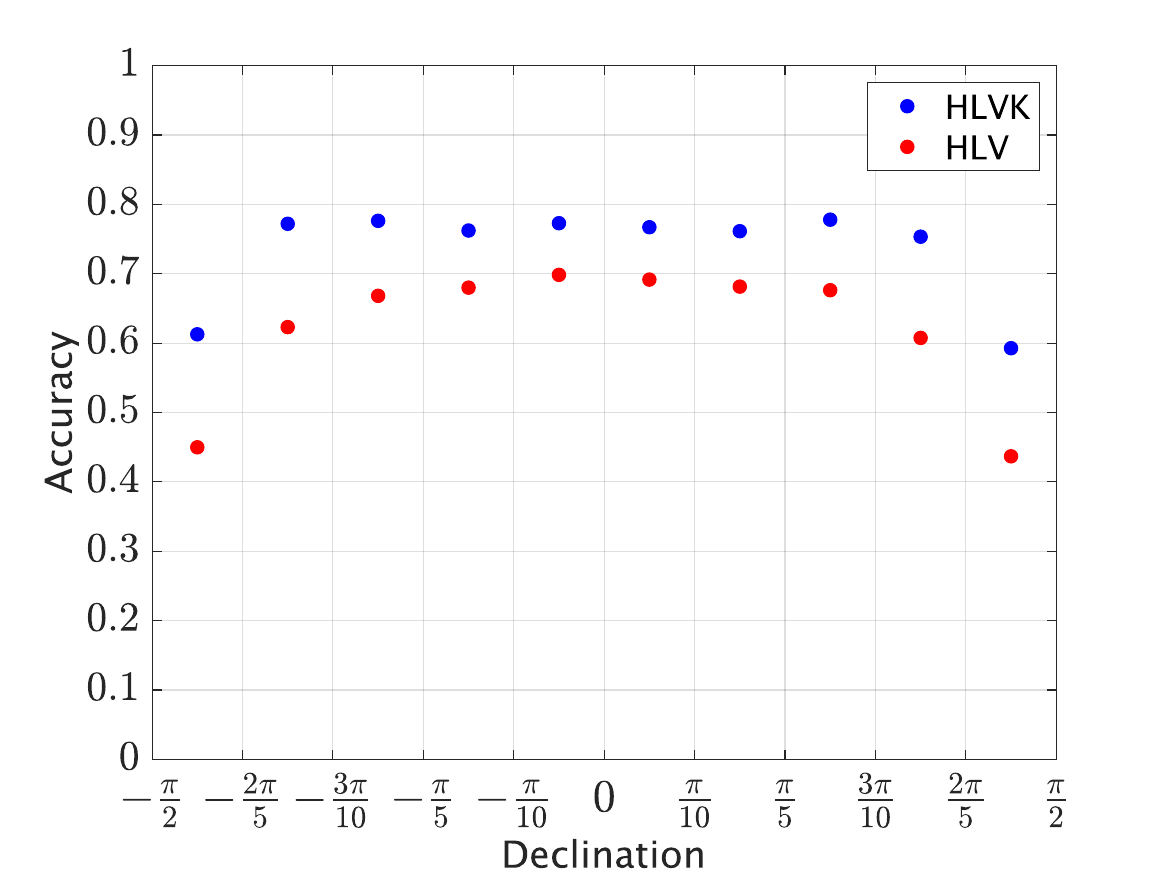}
      \caption{\label{fig:acc_dec} The localization accuracy for 200 sectors vs declination.} 
      \end{minipage} 
    \end{tabular}
  \end{figure*}

The localization accuracy for our test data obtained by methods I, II, and III using the three telescopes is shown in Fig.~\ref{fig:method123}. The accuracy of method II was slightly higher than method I. In method III, the combination of ANN and TCN showed the best accuracy of the three methods for each value of the number of sectors used.

The localization accuracy of method III using three and four telescopes is shown in Fig.~\ref{fig:HLVK}. As shown, the fourth telescope improved the accuracy, and the improvement increased as the number of sectors increased. We also examined the relationship between accuracy and SNR when the number of sectors was 200, as shown in Fig.~\ref{fig:acc_snr}. Our method did not provide good accuracy for a strain with a low SNR, whose noise was much larger than that of the waveform. Figure~\ref{fig:acc_dec} shows the localization accuracy for 200 sectors versus declination. Because declination and right ascension were divided by the same angle, the area of a sector with a high declination was smaller than one with a low declination. Therefore, as shown in Fig.~\ref{fig:acc_dec}, the localization accuracy was low, but the improvement with KAGRA was more significant for GWs from a sector with a high declination. 

The computational cost of classifying a GW signal into one of the 200 sectors by each method is summarized in Table II. This test was performed using an AMD Ryzen 7 3700X processor.

\begin{table}[tb]
    \caption{Time taken by each method} \label{tab:time}
    \centering
    \begin{tabular}{lc}
    \hline
        Method I: ANN & 0.0053 s\\
        Method II: TCN & 0.2652 s\\
        Method III: Combined & 0.2771 s\\
    \hline
    \end{tabular}
\end{table}

%
%


\section{Conclusions\label{sec:conclusions}}
We considered ANN, TCN, and the combination of ANN and TCN to classify GW signals for sky localization. We tested them using our test data from LIGO H1, LIGO L1, and Virgo, and found that the combination of ANN and TCN provided the highest localization accuracy of the three methods. We then trained the model using the four telescopes and compared the localization accuracy with the accuracy using three telescopes. The results showed that the fourth telescope improved the localization accuracy, especially when the number of sectors was large. We also examined accuracy vs. SNR and accuracy vs. declination and found that our model was not effective for strains with low SNR or high declination. To improve the localization accuracy, specifically for data with low SNR, in the future, we intend to apply deep learning algorithms for denoising~\cite{Wei2020, Chatterjee2021} before using our localization method. In future work, we will divide the sky and compare the results in several different ways, including the HEALPix, which is a standard scheme in the electromagnetic counterpart search~\cite{PhysRevD.93.024013}.


\begin{acknowledgments}
We thank Yuting Liu and Kazuki Sakai for their contributions in the early phase of this work and for the enlightening discussions.
This work was supported in part by the Japan Society for the Promotion of Science (JSPS) Grants-in-Aid for Scientific Research on Innovative Areas (Grant No.~24103005, 17H06358) and JSPS Grant-in-Aid for Scientific Research (B) (Grant No.~19H0190).
This work was also supported in part by the Inter-University Research Program of the Institute for Cosmic Ray Research, University of Tokyo, Japan. We would like to thank Editage (www.editage.com) for English language editing.

\end{acknowledgments}

\bibliography{apssamp}

\end{document}